\documentclass[%
 aps,
 prl,
 showkeys,
 amsmath,amssymb,floatfix,
reprint,%
]{revtex4-1}

\usepackage{graphicx}
\usepackage{dcolumn}
\usepackage{bm}
\usepackage{color}

\graphicspath{{figures/}}

\usepackage{hyperref}
\usepackage{cancel}

\usepackage{multirow}
\hypersetup{
   colorlinks,
   linkcolor={blue},
   citecolor={blue},
   urlcolor={blue} 
}

\usepackage{comment}
\usepackage[dvipsnames]{xcolor}  

\begin{document}


\title[
]{\textit{s}+i\textit{f} pairing in Ising superconductors
}

\author{David M\"{o}ckli
}
\email[E-mail me at: ]{d.mockli@gmail.com}
\affiliation{
The Racah Institute of Physics, The Hebrew University of Jerusalem, Jerusalem 9190401, Israel} 
\author{Maxim Khodas} 
\affiliation{
The Racah Institute of Physics, The Hebrew University of Jerusalem, Jerusalem 9190401, Israel}

\date{\today}

\begin{abstract}

We show that an in-plane Zeeman field applied to non-centrosymmetric Ising superconductors converts singlet $s$-wave Cooper pairs to equal-spin triplet $if$ pairs, leading to an enhancement of the critical transition line beyond expected from Ising spin-orbit coupling. 
Singlet to triplet conversion relates to a phase transformation due to spin rotation by the Zeeman field and has a geometric origin.
The discussion is especially relevant, but not limited to monolayer transition metal dichalcogenides. 
\end{abstract}

\maketitle

\textit{Introduction.---}
In non-centrosymmetric superconductors, the presence of momentum odd spin-orbit coupling (SOC) leads to parity-mixed Cooper pair wave functions \cite{Gorkov2001,Bauer2012,Yip2014a}. 
The lack of an inversion center allows for the coexistence of a parity-even singlet and a parity-odd triplet pairing \cite{Sigrist1991}. 
A Zeeman field and SOC affect singlet and triplet Cooper pairs in distinct ways. 
The Zeeman field breaks singlets, which is referred to as paramagnetic limiting. 
This is different for triplets, which might align their spin along the magnetic field avoiding paramagnetic limiting \cite{Frigeri2004,Ramires2018,Fischer2018}.
By contrast, SOC suppresses the equal-spin-triplets ($S_z=\pm 1$), which are (anti) aligned with the effective SOC magnetic field.

The response of a non-centrosymmetric superconductor to a Zeeman field is sensitive to the degree in which singlets and triplets mix. 
At zero Zeeman field, the triplets have zero spin component, $S_z=0$ along the effective SOC magnetic field.
The mixing of such triplets is determined by the ratio of the SOC splitting $\Delta_\mathrm{so}$ and the Fermi energy $E_\mathrm{F}$ \cite{Frigeri2004,Frigeri2006}.
In many cases, $\Delta_\mathrm{so}/E_\mathrm{F}\ll 1$, and the singlets and triplets decouple.

In recent experiments on transition metal dichalcogenides (TMDs), the SOC dramatically enhances the in-plane critical Zeeman field exceeding the Pauli limit \cite{Xi2015,Ugeda2015,Saito2016,Dvir2017,Sohn2018,Nakata2018,DelaBarrera2018}. 
The strong ``Ising'' SOC locks the spins out of plane, and counteracts an in-plane Zeeman field.
The magnetic field-temperature $(B,T)$ phase diagram in the clean limit was obtained in Refs. \cite{Bulaevskii976,Frigeri2004}.
Refs. \cite{Bulaevskii976,Sosenko2017,Ilic2017} showed that the inter-valley impurity scattering suppresses the critical Zeeman field.
Later in Ref.~\cite{Mockli2018}, we showed that the orthogonality of the orbital wave-functions blocks the short-range scattering and stabilizes the critical Zeeman field against the disorder.
The above works considered the singlet pairing interaction.
Indeed, for $\Delta_\mathrm{so}/E_\mathrm{F}\ll 1$, the pairing of $S_z=0$ triplets does not modify the critical transition line $B_\mathrm{c}(T)$ as normally the singlet interactions dominate.

In this work, we show that the Zeeman field converts singlets ($S=0$) into equal-spin-triplets ($S_z=\pm 1$). 
While the Zeeman field induces time-reversal breaking equal-spin-triplets, the SOC promotes singlets.
The competition between Zeeman field and SOC does not depend on the ratio $\Delta_\mathrm{so}/E_\mathrm{F}$ and is controlled by the ratio $B/\Delta_\mathrm{so}$.
According to the previous studies, $B_\mathrm{c}$ is comparable to $\Delta_\mathrm{so}$.
Therefore, $B/\Delta_\mathrm{so} \simeq 1$ at the transition line.
The Cooper pair wave-function, in this case, acquires a substantial triplet component.  
As a result, even a weak interaction of electrons forming equal-spin-triplets has a strong effect on the $(B,T)$ phase diagram. 
Weak attraction or repulsion in the triplet channel leads to enhancement or suppression of $B_\mathrm{c}(T)$ respectively.
This result is crucial for the interpretation of experimental data.


The coexisting order parameters, $S=0$ singlet, $S_z=0$ triplet and $S_z=\pm 1$ triplet transform differently under the crystal symmetry operations. The singlets are scalars of $s$-wave symmetry.
The orbital component of triplet order parameters change sign under the rotation by $\pi/3$, and have $f$-wave symmetry.
The $S_z=0$ triplet order parameter is present even in the absence of Zeeman field and respects the time-reversal symmetry.
The $S_z=\pm 1$ triplets are induced by the Zeeman field and break the time-reversal symmetry of the superconducting state.
For this reason, the order parameter describing these triplets is purely imaginary. 
Here, we study the interplay of Zeeman field and SOC, and mainly focus on real singlet and $S_z=\pm 1$ imaginary triplet order parameters.
The resulting superconducting state has, therefore, $s+i f$ symmetry.

The findings presented here are relevant for non-centrosymmetric superconductors, where the applied magnetic field has an orthogonal component to the effective SOC field. This applies to a large class of materials \cite{Smidman2017}, which besides monolayer TMDs include interface superconductivity \cite{Liu2018} and artificial heterostructures \cite{Shimozawa2016}.

\textit{The Hamiltonian and free energy.---} 
The standard model of a superconductor with anti-symmetric SOC $\boldsymbol{\gamma}_\mathbf{k}=-\boldsymbol{\gamma}_{-\mathbf{k}}$ and a Zeeman field $\mathbf{B}$ is \cite{Bauer2012}
\begin{align}
H & = \sum_{\mathbf{k},s}\xi_\mathbf{k}c^\dag_{\mathbf{k}s}c_{\mathbf{k}s}
+\sum_{\mathbf{k},ss'}\left(\boldsymbol{\gamma}_\mathbf{k}-\mathbf{B} \right )\cdot\boldsymbol{\sigma}_{ss'}c^\dag_{\mathbf{k}s}c_{\mathbf{k}s'} \label{eq:hamiltonian} \\
& +\frac{1}{2}\sum_{\mathbf{k},\mathbf{k}'}\sum_{\{s_i\}}V_{s_1s_2,s_1's_2'}\left(\mathbf{k},\mathbf{k}' \right )c^\dag_{\mathbf{k}s_1}c^\dag_{-\mathbf{k}s_2}c_{-\mathbf{k}'s_2'}c_{\mathbf{k}'s_1'}. \notag
\end{align}
The normal state dispersion $\xi_\mathbf{k}=\xi_{-\mathbf{k}}$ includes the chemical potential.
We define the average over the Fermi surface $\langle|\boldsymbol{\gamma}_\mathbf{k}|^2\rangle_\mathrm{FS}=\Delta_\mathrm{so}^2$.
We use units where $\mathbf{B}$ absorbs usual the prefactor with the $g$-factor and the Bohr magneton $g\mu_\mathrm{B}/2$. 
The interaction in the Cooper channel can be separated into singlet and triplet parts as
\begin{align}
V_{s_1s_2,s_1's_2'}&\left(\mathbf{k},\mathbf{k}' \right )  = \sum_{\Gamma,j} (-v_{s,\Gamma})\left[ \hat{\tau}_{\mathbf{k},\Gamma_j}\right]_{s_1s_2}\left[\hat{\tau}_{\mathbf{k}',\Gamma_j} \right ]^*_{s_1's_2'}\notag  \\
& +\sum_{\Gamma,j} (-v_{t,\Gamma}) \left[\hat{\boldsymbol{\tau}}_{\mathbf{k},\Gamma_j} \right ]_{s_1s_2}
\left[\hat{\boldsymbol{\tau}}_{\mathbf{k}',\Gamma_j} \right ]^*_{s_1's_2'}, 
\label{eq:int}
\end{align}
where $\hat{\tau}_{\mathbf{k},\Gamma_j}=\hat{\psi}_{\mathbf{k},\Gamma_j}i\sigma_y $ and $\hat{\boldsymbol{\tau}}_{\mathbf{k},\Gamma_j}=\hat{\mathbf{d}}_{\mathbf{k},\Gamma_j}\cdot\boldsymbol{\sigma}i\sigma_y$. $j$ labels the basis functions of an irreducible representation $\Gamma$, and $v_{s(t),\Gamma}$ are interactions in each channel and can be attractive (positive) or repulsive (negative).
In a non-centrosymmetric material, 
singlet and triplet channels may belong to the same $\Gamma$ and therefore are allowed to couple \cite{Frigeri2006}.
We 
do not include such 
terms in Eq. \eqref{eq:int}, since as we demonstrate,
parity-mixing is induced primarily by the Zeeman field 
and does not depend on interaction channel mixing. 

We introduce the superconducting mean fields 
$\Delta_{s_1s_2}(\mathbf{k})=\sum_{\mathbf{k}',s_1's_2'} V_{s_1s_2,s_1's_2'}(\mathbf{k},\mathbf{k}')\langle c_{-\mathbf{k}'s_2'} c_{\mathbf{k}'s_1'}\rangle $
that are matrix elements of the gap matrix in spin-space $\Delta_\mathbf{k}=(\psi_\mathbf{k}\sigma_0+\mathbf{d}_\mathbf{k}\cdot\boldsymbol{\sigma})i\sigma_y$. 
The even order parameter $\psi_\mathbf{k}=\psi_{-\mathbf{k}}$ parametrizes singlets, and the odd $d$-vector $\mathbf{d}_\mathbf{k}=-\mathbf{d}_{-\mathbf{k}}$ parametrizes triplets.

We use a path integral approach to obtain 
free energy (See Supplemental Material at [URL will be inserted by publisher] for a detailed derivation)

\begin{align}
F& = - 
\frac{1}{2}\sum_{k,k',s_i}
\Delta^*_{s'_1,s'_2}(\mathbf{k}') 
V^{-1}_{s'_1s'_2,s_1s_2}(\mathbf{k}',\mathbf{k}) \Delta_{s_1,s_2}(\mathbf{k}) 
\label{eq:series}\\
&
+T\sum_{\mathbf{k},\omega_n}\sum_{l=1}^\infty \frac{(-2)^ l}{2l} \mathrm{tr}\left[G(\mathbf{k},\omega_n)\Delta_\mathbf{k} G^\mathrm{T}(-\mathbf{k},-\omega_n)\Delta_\mathbf{k}^\dag\right ]^l \notag,
\end{align}
where $\omega_n=(2n+1)\pi T$ ($k_\mathrm{B}=1$) are Matsubara frequencies, and the normal state Green's function $G(\mathbf{k},\omega_n)$ can be expressed in terms of it's band projections 
\begin{align}
G(\mathbf{k},\omega_n) & = G_+(\mathbf{k},\omega_n)\sigma_0 + G_-(\mathbf{k},\omega_n)\, \mathfrak{g}_\mathbf{k}\cdot\boldsymbol{\sigma}; \\
G_\pm(\mathbf{k},\omega_n) & =\frac{1}{2}\left[\frac{1}{i\omega_n-\epsilon_{\mathbf{k},+}}\pm \frac{1}{i\omega_n-\epsilon_{\mathbf{k},-}}\right],
\label{eq:gpm}\end{align}
where $\epsilon_{\mathbf{k},\pm}=\xi_\mathbf{k}\pm |\boldsymbol{\gamma}_\mathbf{k}-\mathbf{B}|$ and $\mathfrak{g}_\mathbf{k} = (\boldsymbol{\gamma}_\mathbf{k}-\mathbf{B})/|\boldsymbol{\gamma}_\mathbf{k}-\mathbf{B}|$.

\textit{Parity-mixing by Zeeman field.---}
The truncation to quadratic order ($l=1$) in the order parameters of Eq. \eqref{eq:series} determines the
transition line $B_c(T)$.
We introduce the short notation for the products $G_aG_b\equiv G_a(\mathbf{k},\omega_n)G_b(-\mathbf{k},-\omega_n)$ with $a,b=\pm$. Choosing real $\psi_\mathbf{k}$, we calculate the trace in Eq. \eqref{eq:series} for $l=1$
\begin{align}
& \frac{1}{2}\mathrm{tr}\left[G(\mathbf{k},\omega_n)\Delta_\mathbf{k} G^\mathrm{T}(-\mathbf{k},-\omega_n)\Delta_\mathbf{k}^\dag\right ] = G_+G_+\bigr(|\psi_\mathbf{k}|^2+\notag \\ 
& |\mathbf{d}_\mathbf{k}|^2 \bigr )-G_-G_-\bigr[|\psi_\mathbf{k}|^2 \mathfrak{g}_\mathbf{k}\cdot\mathfrak{g}_{-\mathbf{k}}+(\mathfrak{g}_\mathbf{k}\cdot\mathbf{d}_\mathbf{k})(\mathfrak{g}_{-\mathbf{k}}\cdot\mathbf{d}_\mathbf{k}^*) \notag  \\
& -(\mathfrak{g}_{\mathbf{k}}\times\mathbf{d}_\mathbf{k})\cdot(\mathfrak{g}_{-\mathbf{k}}\times\mathbf{d}_\mathbf{k}^*)+2\psi_\mathbf{k}(\mathfrak{g}_\mathbf{k}\times\mathfrak{g}_{-\mathbf{k}})\cdot \mathrm{Im}\,\mathbf{d}_\mathbf{k}\bigr] \notag \\
& -G_+G_-(2\psi_\mathbf{k}\,\mathfrak{g}_{-\mathbf{k}}\cdot\mathrm{Re}\,\mathbf{d}_\mathbf{k}-\mathfrak{g}_{-\mathbf{k}}\cdot\mathbf{q}_\mathbf{k}) \notag \\
&+G_-G_+(2\psi_\mathbf{k}\,\mathfrak{g}_{\mathbf{k}}\cdot\mathrm{Re}\,\mathbf{d}_\mathbf{k}+\mathfrak{g}_{\mathbf{k}}\cdot\mathbf{q}_\mathbf{k}),
\label{eq:trace}
\end{align}
where $\mathbf{q}_\mathbf{k}=i\mathbf{d}_\mathbf{k}\times\mathbf{d}_\mathbf{k}^*$. 
If the superconducting states respect time-reversal symmetry ($\mathrm{Im}\,\mathbf{d}_\mathbf{k}=0$), the singlet-triplet mixing occurs in $a\neq b$ terms only. 
Such terms, however, are proportional to the difference of density of states of the two Fermi sheets at the Fermi level $E_\mathrm{F}$, $\Delta N=N_+-N_-$, which gives a contribution of the order $\Delta_\mathrm{so}/E_\mathrm{F}$ \cite{Frigeri2004}. 
Then, if $\Delta_\mathrm{so}/E_\mathrm{F}\ll 1$, the singlet and triplet channels decouple at the quadratic level and can be studied separately. 
As the Zeeman field breaks time-reversal, singlet-triplet coupling arises via the term of Eq.~\eqref{eq:trace} proportional to  $2\psi_\mathbf{k}(\mathfrak{g}_\mathbf{k}\times\mathfrak{g}_{-\mathbf{k}})\cdot \mathrm{Im}\,\mathbf{d}_\mathbf{k}$, which is non-negligible even when $\Delta_\mathrm{so}/E_\mathrm{F}\ll 1$.

\begin{table}
\caption{Cooper channels of $D_{3h}$ with its even-singlet and odd-triplet basis functions. Here we use 
$\hat{\gamma}_\mathbf{k}\propto -\sin k_x+2\sin\left(k_x/2 \right ) \cos\left(k_y\sqrt{3}/2 \right )$.
}
\label{tab:irreps}
\begin{tabular}{ccccc}
\hline
Irrep & Singlet $\hat{\psi}_\mathbf{k}$ & Triplet $\hat{\mathbf{d}}_\mathbf{k}$ & Order par. & Limited by \\ \hline
$A_1'$ ($s$) & $1$ &  & $\psi_0$ & Zeeman field \\
$A_1'$ ($f$) &  & $\hat{\gamma}_\mathbf{k}\hat{\boldsymbol{z}}$ & $\eta_z$ & -- \\
$E''$ ($if$) &  & $\hat{\gamma}_\mathbf{k}\hat{\boldsymbol{x}};\hat{\gamma}_\mathbf{k}\hat{\boldsymbol{y}}$ & $\eta_x,\eta_y$ & Ising SOC \\ \hline
\end{tabular}
\end{table}

\begin{figure*} 
\centering
\includegraphics[width=0.98\textwidth]{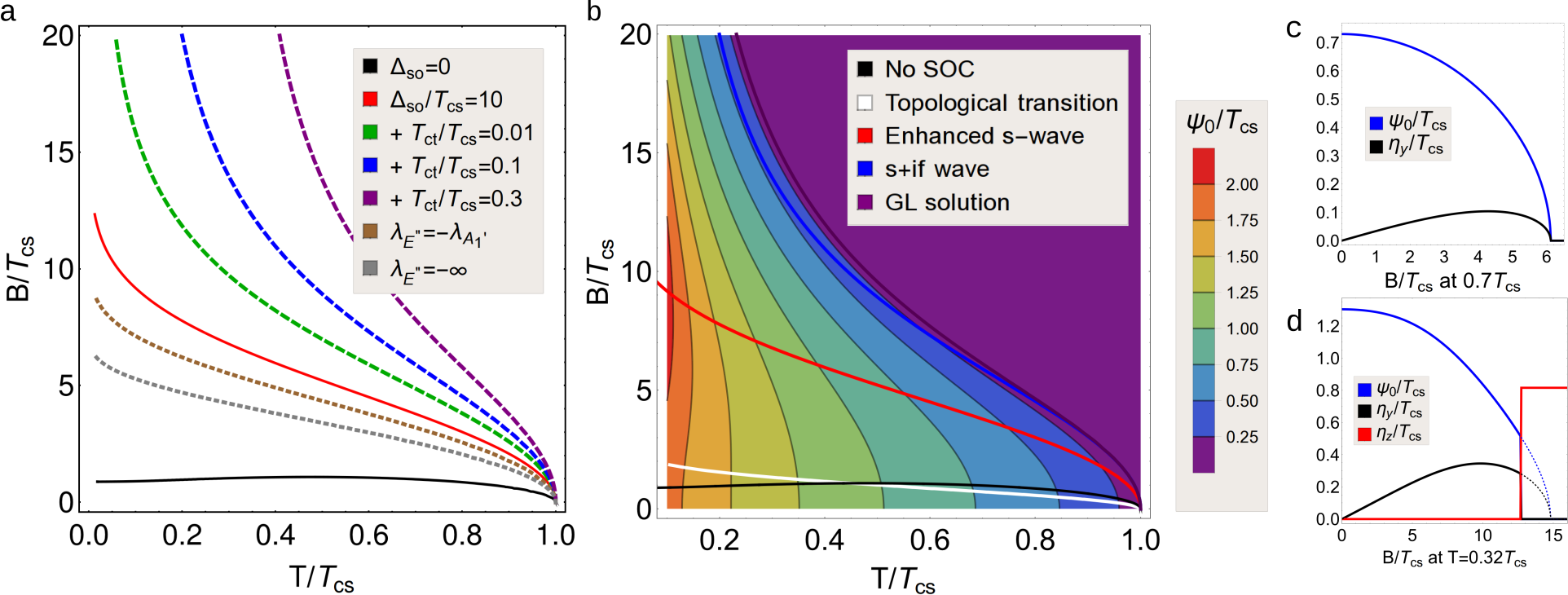}
\caption{
\label{fig:one} 
(a) $B_\mathrm{c}(T)$ obtained from the pair breaking equation $\alpha_s(T,B)\alpha_t(T,B) = \alpha_{st}^2(B)$. The brown and the gray dotted curves show repulsion in the $E''$ channel, with the dimensionless couplings $\lambda_{A_1'}=N_0v_{s,A_1'}$ and $\lambda_{E''}=N_0v_{t,E''}$. 
(b) GL solution for $\psi_0$ with $T_\mathrm{ct}/T_\mathrm{cs}=0.1$. The nodal transition line is determined by the condition $B=\psi_0$. 
The black, red and blue curves are the same as in (a). The blue and purple transitions deviate far from $T_\mathrm{cs}$, where the GL solution fails, but still provides a qualitative description.
(c-d) Plots of $\psi_0(B)$, $\eta_y(B)$ and $\eta_z(B)$ at two different temperatures. $\eta_z$ is favourable at high magnetic fields below its own critical temperature, in this case $T_\mathrm{ctz}=T_\mathrm{cs}/2$ for illustrative purposes.
}
\end{figure*}

\begin{figure} 
\centering
\includegraphics[width=0.48\textwidth]{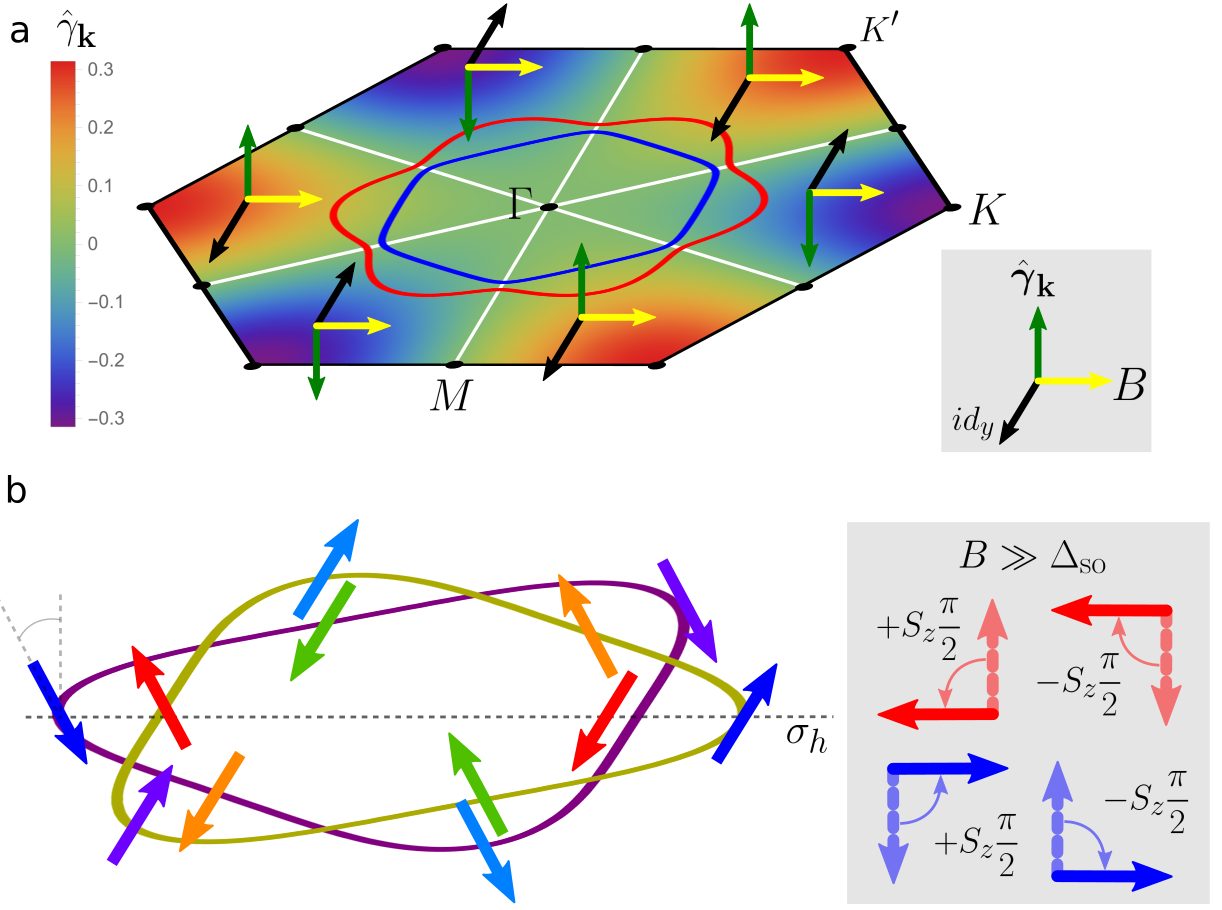}
\caption{\label{fig:two} 
(a)
Illustration of the Brillouin zone, showing the sign modulation of SOC and parallelepipeds $\mathbf{B}\times\boldsymbol{\gamma}_\mathbf{k}\cdot i\mathbf{d}_\mathbf{k}$. 
(b)
Schematic of the Cooper pairs nested on the spin-split Fermi-surface. Cooper pairs are related by the symmetry $\sigma_h\mathcal{T}\nearrow=\sigma_h\swarrow=\searrow$ and are shown as same color arrows.
}
\end{figure}

\textit{Ising superconductors.---}
To work out a concrete example, we consider monolayer TMDs with point group symmetry $D_{3h}$.
We assume $\Delta_\mathrm{so}/E_\mathrm{F}\ll 1$, which allows us to neglect the $a\neq b$ terms in Eq. \eqref{eq:trace}, and write $\sum_\mathbf{k}\rightarrow N_0 \int_0^{2\pi}\frac{\mathrm{d}\varphi}{2\pi}\int_{-\epsilon_c}^{\epsilon_c}\mathrm{d}\xi$, where $N_0$ is the density of states at the Fermi level of the two Fermi sheets and $\epsilon_c$ is a characteristic cutoff energy of the pairing interaction. 
We specialize to the case where $\mathbf{B}\perp \boldsymbol{\gamma}_\mathbf{k}\parallel\hat{\boldsymbol{z}}$, relevant for Ising superconductors in general \cite{Mockli2018}. 
For this special case, the band splittings at opposite momenta $\mathbf{k}$ remain the same $|\boldsymbol{\gamma}_{\pm\mathbf{k}}-\mathbf{B}| = \sqrt{|\boldsymbol{\gamma}_\mathbf{k}|^2+B^2}$, ensuring perfect Fermi surface nesting for Cooper pairing; see Fig. (\ref{fig:two}). 
Then, singlets and triplets only mix in the triple product
\begin{align}
    2\psi_\mathbf{k}(\mathfrak{g}_\mathbf{k}\times\mathfrak{g}_{-\mathbf{k}})\cdot \mathrm{Im}\,\mathbf{d}_\mathbf{k}\propto \mathbf{B}\times\boldsymbol{\gamma}_\mathbf{k}\cdot\mathrm{Im}\,\mathbf{d}_\mathbf{k}.
    \label{eq:triple}
\end{align}
Only imaginary in-plane components of the $d$-vector breaking time-reversal contribute to the triple product.

We expand the singlet and triplet order parameters of a specific $l_\Gamma$-dimensional irreducible representation $\Gamma$ of $D_{3h}$ in terms of hatted basis functions as
$\psi_{\mathbf{k},\Gamma}=\sum_{i=1}^{l_\Gamma}\psi_{\Gamma_i}\hat{\psi}_{\mathbf{k},\Gamma_i}$ and $\mathbf{d}_{\mathbf{k},\Gamma}=\sum_{i=1}^{l_\Gamma}\eta_{\Gamma_i}\hat{\mathbf{d}}_{\mathbf{k},\Gamma_i}$, where $\psi_{\Gamma_i}$ and $\eta_{\Gamma_i}$ serve as complex Ginzburg-Landau (GL) order parameters of the singlet and triplet component, respectively. 
We consider the singlet channel $\psi_{\mathbf{k},A_1'}=\psi_0 1$, and two channels for the triplets: $\mathbf{d}_{\mathbf{k},A_1'}=\eta_z\hat{\boldsymbol{\gamma}}_\mathbf{k}$, where $\hat{\boldsymbol{\gamma}}_\mathbf{k}=\boldsymbol{\gamma}_\mathbf{k}/|\boldsymbol{\gamma}_\mathbf{k}|$ with $\boldsymbol{\gamma}_\mathbf{k}=\gamma_\mathbf{k}\hat{\boldsymbol{z}}$, and $\mathbf{d}_{\mathbf{k},E''}=\eta_x \hat{\gamma}_\mathbf{k}\hat{\boldsymbol{x}}+\eta_y \hat{\gamma}_\mathbf{k}\hat{\boldsymbol{y}}$; see table \ref{tab:irreps}.
With this decomposition into channels, we can write $\psi_\mathbf{k}=\psi_0$ and $\mathbf{d}_\mathbf{k}=\hat{\gamma}_\mathbf{k}(\eta_x,\eta_y,\eta_z)$, keeping in mind that $\{\psi_0,\eta_z\}$ belong to $A_1'$ and $\{\eta_x,\eta_y\}$ to $E''$.

The triple product in Eq. \eqref{eq:triple} mixes the $A_1'$ singlet ($s$-wave) and the $E''$ triplet ($if$-wave) channels. The resultant parity-mixed superconducting state is referred to as $s+if$. 
Therefore, because of the interaction in the $E''$ channel, the Zeeman field induces equal-spin triplets. Without loss of generality, we can fix the direction of $\mathbf{B}=B\hat{\boldsymbol{x}}$, such that we rewrite $\mathbf{d}_\mathbf{k}=\hat{\gamma}_\mathbf{k}(0,i\eta_y,\eta_z)$, where $\eta_y$ and $\eta_z$ are now real. 
In the limit $\Delta_\mathrm{so}/E_\mathrm{F}\ll 1$, $\eta_z$ decouples from the $\{\psi_0,\eta_y\}$ subsystem at the quadratic level. The decoupled triplet $\mathbf{d}_{\mathbf{k},A_1'}\parallel \boldsymbol{\gamma}_\mathbf{k}$ is protected from both the Zeeman field and SOC, and for this reason, we focus on the $\{\psi_0,\eta_y\}$ subsystem. 
The main point is: although the basis functions of $A_1'$ and $E''$ are orthogonal, they mix due to the Zeeman field in Eq. \eqref{eq:triple}.

\textit{The $B_\mathrm{c}(T)$ transition.---}
We now obtain the continuous superconducting to normal state transition lines $B_\mathrm{c}(T)$ using Eq. \eqref{eq:series} with $l=1$. 
The energy integrals followed by the Matsubara summation can be performed to obtain (See Supplemental Material at [URL will be inserted by publisher] for details)
\begin{align}
& T\sum_{\omega_n}\int\mathrm{d}\xi G_+G_+  =\log\left[2e^\gamma\epsilon_c/(\pi T)\right]-C(\rho_\mathbf{k})/2;\label{eq:gpgp}\\
& T\sum_{\omega_n}\int\mathrm{d}\xi \,G_-G_-  =C(\rho_\mathbf{k})/2;\label{eq:gmgm}\\
& C(\rho_\mathbf{k})  = \mathrm{Re}\,\psi\left(\frac{1}{2}+i\frac{\rho_\mathbf{k}}{2}\right )-\psi\left(\frac{1}{2}\right)\geq 0, \label{eq:digamma}
\end{align}
where $\psi(z)$ is the digamma function, $\rho_\mathbf{k} = \sqrt{|\boldsymbol{\gamma}_\mathbf{k}|^2 +B^2}/(\pi T)$, and $\gamma$ is Euler's constant. 

From Eqs. (\ref{eq:gpgp},\ref{eq:gmgm},\ref{eq:digamma}) and the trace \eqref{eq:trace}, we obtain the quadratic free energy
\begin{align}
\frac{1}{2N_0}F_{T,B}^{(l=1)}\left[\psi_0,\eta_y\right] & = \alpha_{s}(T,B)\psi_0^2 + \alpha_t(T,B)\eta_y^2 \notag \\
& + 2\alpha_{st}(B)\psi_0\eta_y,
\label{eq:free_quadratic}
\end{align}
with the coefficients defined as
\begin{subequations}
\label{eq:alpha}
\begin{align}
\alpha_s(T,B) & =\ln\left(\frac{T}{T_\mathrm{cs}}\right)+C(\rho)\frac{B^2}{\Delta_\mathrm{so}^2+B^2}; \label{eq:as} \\
\alpha_t(T,B) & = \ln\left(\frac{T}{T_\mathrm{ct}}\right)+C(\rho)\frac{\Delta_\mathrm{so}^2}{\Delta_\mathrm{so}^2+B^2}; \label{eq:at} \\
\alpha_{st}(B) & = -C(\rho)\frac{B\Delta_\mathrm{so}}{\Delta_\mathrm{so}^2+B^2}.
\label{eq:ast} 
\end{align}
\end{subequations}
$T_\mathrm{cs} (T_\mathrm{ct})$ is the singlet (triplet) critical transition temperature determined by $(N_0 v_{s,A_1'})^{-1}=\ln\left[2 e^\gamma\epsilon_c /(\pi T_\mathrm{cs}) \right ]$ and $(N_0 v_{t,E''})^{-1}=\ln\left[2 e^\gamma\epsilon_c /(\pi T_\mathrm{ct}) \right ]$, and $\rho = \sqrt{\Delta_\mathrm{so}^2 +B^2}/(\pi T)$. 
Eq. \eqref{eq:free_quadratic} clearly shows the limiting mechanisms acting on the singlet and triplet components. 
Positive terms in Eq. \eqref{eq:free_quadratic} suppress the superconducting state and negative terms stabilize it. 
$\alpha_s$ shows that the Zeeman field limits the $s$-wave singlets $\psi_0$.
By contrast, $\alpha_t$ shows that SOC limits the $if$-wave triplets $\eta_y$. 
With different limiting mechanisms (Zeeman field and SOC) affecting different order parameters ($\psi_0$ and $\eta_y$), their mixing via $\alpha_{st}$ can be interpreted as a conversion of $s$-wave singlets to equal-spin $if$-wave triplet Cooper pairs by the Zeeman field.
Interestingly, $\alpha_{st}(T,B)$ vanishes in purely triplet superconductors, where anti-symmetric SOC vanishes.

Minimization of the free energy Eq. \eqref{eq:free_quadratic} yields the pair-breaking equation
$\alpha_s(T,B)\alpha_t(T,B) = \alpha_{st}^2(B)$ that 
determines $B_\mathrm{c}(T)$. 
If the attraction exists only in the $s$-wave singlet channel, the above condition reduces to the pair-breaking equation $\alpha_s(T,B)=0$, which is found in Refs. \cite{Frigeri2004,Ilic2017,Liu2018}.
We plot $B_\mathrm{c}(T)$ in Fig. \ref{fig:one}a, which is very sensitive to $if$ components.

\textit{The Ginzburg-Landau (GL) regime.---}
To obtain the order parameters $\{\psi_0,\eta_y\}$ 
in the superconducting phase, we keep the quartic terms $(l=2)$ in the GL expansion \eqref{eq:series} near $T=T_\mathrm{cs}$,
\begin{align}
&\frac{1}{2N_0}F_{T,B}[\psi_0,\eta_y] = \alpha_s(T_\mathrm{cs},B)\psi_0^2+ \alpha_t(T_\mathrm{cs},B)\eta_y^2 \notag \\
& +2\alpha_{st}(B)\psi_0\eta_y +\beta_1(B)\left(B\psi_0-\Delta_\mathrm{so}\eta_y\right)^2\left(\Delta_\mathrm{so}\psi_0+B\eta_y\right)^2 \notag \\
& +\beta_2(B)\left(B\psi_0-\Delta_\mathrm{so}\eta_y\right)^4+\beta_3(B)\left(\Delta_\mathrm{so}\psi_0+B\eta_y \right )^4,
\label{eq:free_4}
\end{align}
where the coefficients of the quadratic terms are defined in Eq.~\eqref{eq:alpha} and the coefficients of the quartic terms are
\begin{subequations}
\begin{align}
\beta_1(B) & = \frac{\mathrm{Im}\,\psi^{(1)}\left(\frac{1}{2}-i\frac{\rho_\mathrm{cs}}{2} \right )}{2\pi T_\mathrm{cs}\left(\Delta_\mathrm{so}^2+B^2 \right )^{5/2}}-\frac{C(\rho_\mathrm{cs})}{\left(\Delta_\mathrm{so}^2+B^2 \right )^{3}};
\label{eq:b1}\\
\beta_2(B) & =-\frac{\mathrm{Re}\,\psi^{(2)}\left(\frac{1}{2}+i\frac{\rho_\mathrm{cs}}{2}; \right )}{16\pi^2T_\mathrm{cs}^2\left(\Delta_\mathrm{so}^2+B^2 \right )^2};
\label{eq:b2}
\\
\beta_3(B) & = \frac{7\zeta(3)}{8\pi^2T_\mathrm{cs}^2\left(\Delta_\mathrm{so}^2+B^2 \right )^2},
\label{eq:b3}
\end{align}
\end{subequations}
with $\rho_\mathrm{cs} = \rho(T_\mathrm{cs})$, and the
poly-gamma functions are defined as $\psi^{(m)}=\frac{\mathrm{d}^m}{\mathrm{d}z^m} \psi(z)$. 
In Fig. \ref{fig:one}(b-d) we show the the order parameters in the superconducting phase obtained by minimizing the free energy, Eq. \eqref{eq:free_4}.

\textit{Discussion and conclusion.---}
The singlet to triplet conversion has a transparent geometrical interpretation, see Fig.~\ref{fig:two}.
At $B=0$, SOC polarizes the electron states out of the plane so that the spin-up and spin-down Fermi lines cross, Fig.~\ref{fig:two}(a,b).
The Cooper pair is singlet formed by the two spin states at momenta $\pm\bm{k}$,
$\Psi_s = |\bm{k},\uparrow; -\bm{k},\downarrow\rangle - |\bm{k},\downarrow;-\bm{k},\uparrow\rangle$, where $|m;n\rangle$ denotes the two-fermion state with $m$ and $n$ being the quantum numbers for each of the occupied states.

Consider the transformation of $\Psi_s$ induced by strong Zeeman field $B \gg \Delta_\mathrm{so}$; Fig.~\ref{fig:two}(b).
In this limit, all the spins are polarized in-plane, so that the original singlet $\Psi_s$ transforms to
(see Fig.~\ref{fig:two}(b), inset) 
\begin{align}\label{eq:t}
\Psi_t = |\bm{k},U_{-\frac{\pi}{2}}^{\hat{y}}\uparrow; -\bm{k},U_{\frac{\pi}{2}}^{\hat{y}}\downarrow\rangle \!-\! |\bm{k},U_{-\frac{\pi}{2}}^{\hat{y}}\downarrow;-\bm{k},U_{\frac{\pi}{2}}^{\hat{y}}\uparrow\rangle  ,
\end{align}
where $U^{\mathbf{n}}_{\varphi}$ is the operator of spinor rotation around an axis along $\mathbf{n}$ by an angle $\varphi$. 
Since $|\downarrow \rangle = | U_{\pi}^{\hat{y}} \uparrow \rangle $ and the geometrical phase
$U_{2\pi}^{\hat{y}} = -1$, Eq.~\eqref{eq:t} reduces to
\begin{align}\label{eq:tt}
\Psi_t = |\bm{k},\uparrow; -\bm{k},\uparrow\rangle \!+\! |\bm{k},\downarrow;-\bm{k},\downarrow\rangle\, .
\end{align}
%
As expected, the Zeeman field converts the singlet $\Psi_s$ into a $S_z = \pm 1$ triplet state Eq.~\eqref{eq:tt}. 
This state is odd under time-reversal $\mathcal{T}$ 
thanks to the geometric phase and is parameterized by $\mathrm{Im} [\mathbf{d}_{\mathbf{k}}]_y \neq 0$. 
Being triplet, it is also odd in $\mathbf{k}$. 
The only combination that satisfies the above requirements is $\mathbf{d}_{\mathbf{k}} \propto i \mathbf{\gamma}_{\mathbf{k}} \times {\mathbf{B}}$, see Fig.~\ref{fig:two}(a).

We now discuss the dependence of the singlet-to-triplet conversion on SOC.
The Zeeman field is effective only if it couples occupied and unoccupied states split by SOC. 
Therefore, when SOC is smaller than the superconducting gap, the spin-triplet conversion is negligible.
In the opposite limit, the phase space available for a converted pairs scales with the spin splitting
$\sqrt{\Delta_\mathrm{so}^2 + B^2}$.
This is the reason for the logarithmic enhancement of the spin-triplet mixing terms in Eq.~\eqref{eq:alpha}, $\alpha_{s,t,st} \propto C(\rho) \approx \ln(\sqrt{B^2+\Delta_\mathrm{so}^2}/T)$ in the strong SOC limit.

As shown in Fig. \ref{fig:one}a, a weak attraction has a strong effect on $B_\mathrm{c}(T)$ while the effect of repulsion is less pronounced.
Close to $T_\mathrm{cs}$, the enhancement of the critical field can be obtained from the pair-breaking equation in the limit $\Delta_\mathrm{so}/T_\mathrm{cs}\gg 1$, which gives
\begin{align}
\frac{B_\mathrm{c}^2(T)}{\Delta_\mathrm{so}^2}=
\left(\frac{\ln\frac{\Delta_\mathrm{so}}{\Delta_\mathrm{ct}}}{\ln\frac{T_\mathrm{cs}}{T_\mathrm{ct}}} \right )\frac{1-\frac{T}{T_\mathrm{cs}}}{\ln\frac{\Delta_\mathrm{so}}{\Delta_\mathrm{cs}}},
\label{eq:bct}
\end{align}
where $\Delta_\mathrm{cs(ct)}=(\pi/2 e^\gamma)T_\mathrm{cs(ct)}$. The term in parentheses gives the enhancement due to the presence of triplets.

According to Eq. \eqref{eq:bct} for $\Delta_\mathrm{ct} \lesssim \Delta_\mathrm{sc}$ the critical field is enhanced by a factor of $\simeq \sqrt{\ln(\Delta_\mathrm{so}/\Delta_\mathrm{ct} ) }$, which can be substantial as in TMDs the SOC may exceed the superconducting gap by more than three order of magnitudes \cite{Dvir2017}.
In summary, the conversion of $s$-wave singlets to $if$ triplets by the Zeeman field is of importance both theoretically and for interpreting the experimental data.

\begin{acknowledgments}
We thank G. Blumberg, T. Dvir, and H. Steinberg for enlightening discussions.
We acknowledge the financial support by the Israel Science Foundation, Grant No. 1287/15 and D.M. also acknowledges the support from the Swiss National Science Foundation, Project No. 184050. 
\end{acknowledgments}


\bibliographystyle{apsrev4-1}
\bibliography{bibliography}

\end{document}